# Mechanisms for QoE optimisation of Video Traffic: A review paper

Qahhar Muhammad Qadir, Alexander A. Kist, and Zhongwei Zhang
University of Southern Queensland
Queensland, Australia
{safeen.qadir,kist}@ieee.org, zhongwei.zhang@usq.edu.au

**Abstract**
Transmission of video traffic over the Internet has grown exponentially in the past few years with no sign of waning. This increasing demand for video services has changed user expectation of quality. Various mechanisms have been proposed to optimise the Quality of Experience (QoE) of end users' video. Studying these approaches are necessary for new methods to be proposed or combination of existing ones to be tailored. We discuss challenges facing the optimisation of QoE for video traffic in this paper. It surveys and classifies these mechanisms based on their functions. The limitation of each of them is identified and future directions are highlighted.

**Index Terms**
*QoE; Video; Quality optimisation*

## I. INTRODUCTION

The video storm has already started (Cisco, 2014) which made studying of Quality of Experience (QoE) inevitably important. QoE is a measure to evaluate the service quality as perceived by end users (ITU-T, 2007). Various technical and non-technical factors affect this new quality measure (Brooks & Hestnes, 2010). Among these factors are those related to service preparation, delivery and presentation. Maintaining QoE at an acceptable level is a challenging task. We will discuss these challenges in Section II.

Extensive research has been done in the area of QoE optimisation for video traffic. Many solutions have been introduced to tackle the challenge of increasing video traffic such as WiFi offloading (Maallawi, Agoulmine, Radier, & ben Meriem, 2014). Mechanisms are also required to meet the satisfaction of users and preserve the interests of service providers. This common goal has been targeted by various designs. Different approaches are available which focus on different optimisation metrics, scope and adaptation methods. They can be deployed individually or jointly to achieve this goal which is called cross-layer design in the later case (Fu, Munaretto, Melia, Sayadi, & Kellerer, 2013).

Optimisation has to address the conflict of interest of both end users and network providers. From end users' perspective, maximum quality is expected whereas low-cost and the number of served users are important from the network providers' perspective. These two can be jointly optimised through an intelligent design. The emerging demand for video quality has also promoted the development of cross-layer designs for video transmission that are QoE-aware. They have been proposed as solutions to address the aforementioned challenge. The main objective is to utilise network resources efficiently through the cooperation between layers and optimisation of their parameters. As a result of this cooperation, a maximum possible quality for as many users as possible is expected.

There have been similar efforts to review and survey mechanisms for QoE optimisation of video traffic (Ernst, Kremer, & Rodrigues, 2014; Maallawi et al., 2014). Most recently, a comprehensive survey is presented in (Maallawi et al., 2014) on the offload approaches at different parts of the global network (access, core, gateway). Offloading is a possible way to optimise QoE and manage resources efficiently. The primary objective is to maintain the perceived QoE by redirecting part of traffic to alternate cost effective paths or enabling direct communication between nearby devices. This frees up costly congested paths for the 3GPP Radio Access Network (RAN) (4G/3G/2G) and Mobile Packet Core Network (MPCN) and avoids transporting low priority traffic on these paths. The survey discusses the alternative paths of offloading and their management in the access and core network. It also compares the offload approaches and raises open issues to be tackled in managing offload such as architecture to adopt, decision making process to design and required information for decisions. Another similar survey was done in (Ernst et al., 2014). Recent mechanisms within the Heterogeneous Wireless Network (HWN)s are categorised according to their functions (handover, MAC and scheduling, topology and power control). A comparison between approaches is made for each category. The limitation of each approach is also explained and potential trends in the area are identified.

This work broadens the offloading survey in (Maallawi et al., 2014) and HWN survey in (Ernst et al., 2014). It is a comprehensive survey which covers all proposed solutions in the area of QoE optimisation for video traffic in the last 10 years independent of a specific solution or underlying network. Thus, it can be a good tutorial for interested readers in this area. The main contributions of this paper are as follow:

1. Categorisation of mechanisms proposed for QoE optimisation of video traffic in the last 10 years,
2. Comparison of various mechanisms of each category, and
3. An outline of future work in the area of QoE optimisation for video traffic.

The paper is structured as follows. Section II discusses the motivation for this work and challenges in the optimisation of QoE. Section III reviews related work in the area of QoE optimisation of video. Section IV presents open issues for future research. The paper is concluded with Section V.

## II. QOE OPTIMISATION CHALLENGES AND MOTIVATIONS

Various media types have different metrics, and thus are hard to be compared. QoE compared to the traditional Quality of Service (QoS) is more complex to be satisfied under highly dynamic environment. This is due to the multidimensional requirements of current services. It is a subjective metric and hard to be quantified. The evolution of video capable devices such as smartphones which can connect to the Internet anywhere anytime, has changed users consumption behaviour from traditional text-based surfing to real-time video streaming. The media and network operators have been challenged by the huge volume of video traffic and users' high expectation of quality. They face a crucial task of maintaining a satisfactory QoE of streaming services (Xu et al., 2014). Non-optimised designs of mobile applications running these devices have wasted expensive radio resources and the limited licensed spectrum at the access level is not in the favour of all required services. To meet users' rising demand for bandwidth, operators need to increase the capacity of their network by deploying more spectrum which is expensive and not always available. For example, in 2011, the French regulator ARCEP attributed 4G/800MHz band in France, where 2.639 billion Euros was estimated for a 30MHz duplex and 0.94 billion of Euros for a 70MHz duplex belonging to the 4G 2.6GHz band. This high demand has initiated the need for upgrading network components which again associates significant additional costs. Instead, operators work around the problem by putting less expensive solutions such as content caching over the top services (e.g. Youtube) inside their Autonomous System (AS) which avoids costly inter AS traffic. Other than the technical challenges, service providers are also facing business challenges. Giant companies such as Google and Apple started to offer services traditionally provided by service providers (Maallawi et al., 2014).

In the last few years, mobile network operators have been losing revenue from the fixed and mobile services (Maallawi et al., 2014). The traditional time-based billing is now obsolete and has been replaced with a monthly-based fix rate regardless of consumed data capacity. In addition to that, users keep switching to cheaper charging providers. This increase in data traffic and decrease in average revenue per user demand new mechanisms which can reduce the operational costs and optimise video transmission (Fu et al., 2013). Simply throwing bandwidth at the problem is not a solution (Roberts, 2009).

The above challenges have motivated researchers and service providers to find better cost-effective solutions. They should be able to optimise the utilisation of resources with the aim to maximise users' satisfaction on delivered services.

## III. OPTIMISATION OF QOE FOR VIDEO TRAFFIC

Whilst there are many studies focused on the optimisation of image and voice, we review only those which targeted the video services. We categorise these approaches based on their functions (rate adaptation, cross-layer mechanisms, scheduling, content and resource management). Specifically, we focus on methods that optimise the QoE of video traffic.

### A. QoE optimisation through rate adaptation

Adaptive video rate is not a new topic, it has been proposed by various authors to enhance the video quality. Work in (Piamrat, Ksentini, Bonnin, & Viho, 2009) proposes online estimation of QoE using a tool called Pseudo Subjective Quality Assessment (PSQA). The video rate is adapted dynamically for multicast communication in wireless LAN employing the tool. The multicast transmission rate is decreased when the user experiences poor QoE and is increased otherwise. Assuming that every multicast node runs PSQA, the multicast data rate is adapted by the access point at the MAC level. The simulation outcome showed that QoE and the wireless channel utilisation are increased compared to existing solutions including the IEEE 802.11 standard. The tool is based on statistic learning using random neural network which is trained to learn mapping between QoE scores and technical parameters. It has to be re-trained whenever a new parameter needs to be taken into consideration. The application of this work is limited to the same wireless LAN where the access points are located.

The user-centric discretised streaming model presented in (Liu, Rosenberg, Simon, & Texier, 2014) is specifically designed for live rate-adaptive streaming in modern Content Delivery Network (CDN). The objectives are to maximise the minimum satisfaction among users and average satisfaction of all users. Algorithms also proposed for the CDN's content placement, content delivery and user assignment. The system with limited CDN infrastructure in a dynamic environment achieved high user satisfaction through a large simulation campaign.

Work in (A. Khan, Sun, Jammeh, & Ifeachor, 2010) utilise a QoE prediction model from their previous work to achieve Sender Bit Rate (SBR) adaptation for video over wireless that is suitable for network resources and content types. For a requested QoE, an appropriate SBR is identified by the content providers and optimised resources are provided by the network operators. The shortcoming of the study is that QoE (on which the rate is adapted) is predicted based on a limited number of parameters such as content type, sender bit-rate and frame-rate from the application layer and packet error ratio from the network layer.

A QoE and proxy based multi-stream scalable (temporal and amplitude) video adaptation for wireless network is presented in (Hu et al., 2012). According to the simulation results, it outperforms the TCP Friendly Rate Control (TFRC) in terms of agility to track link quality, support for differentiated services and fairness with conventional TCP flows. The proxy at the edge of a wireless network maximises the weighted sum of video qualities of different streams by iteratively allocating rate for each stream. This is based on their respective rate-quality relations, wireless link throughputs and the sending buffer status (without feedback from receivers). The subjective quality is related to a given rate by choosing the optimal frame rate and quantisation stepsize through an analytical rate-quality tradeoff model. The study is limited to layered videos and justification needed for quality estimation without the feedback from receivers.

An adaptive streaming scheme presented in (Koo & Chung, 2010) called Mobile-aware Adaptive Rate Control (MARC) adjusts the quality of bit-stream and transmission rate of video streaming in mobile broadband network. It is done based on the status of the wireless channel and network as well as client buffer for Scalable Video Coding (SVC). An Additive-Increase Heuristic-Decrease (AIHD) congestion control is proposed to reduce rate oscillation. Simulation results show that MARC can control the transmission rates of video streaming based on the mobile station status in the wireless network, though its limited to layered videos such as SVC. A comparison of mechanisms relying on adapted rate for QoE optimisation is illustrated in Table I.

Table I. COMPARISON OF RATE ADAPTATION MECHANISMS

| Ref. | Approach | Traffic | Date | Underlying network | QoE measurement | Limitations |
|---|---|---|---|---|---|---|
| (Piamrat et al., 2009) | MOS-based rate adaptation | Video | 2009 | Wireless LAN | PSQA-based MOS | PSQA needs to be re-trained for new QoE parameters |
| (Liu et al., 2014) | Live rate adaptation | Streaming Video | 2014 | ADSL, WiFi, 3G | Utility function dependent on encoding bitrate | Missing subjective MOS |
| (A. Khan et al., 2010) | QoE-driven rate adaptation | Video | 2010 | Wireless | PSNR-MOS mapping | MOS mapped from PSNR |
| (Hu et al., 2012) | Quality-based rate allocation | Video | 2012 | Wireless | Utility function dependent on rate | Normalised quality calculated from proposed rate-quality model |
| (Koo & Chung, 2010) | Transmission rate adaptation | Video | 2010 | Mobile broadband | PSNR | MOS mapped from PSNR |

*B. QoE optimisation through cross-layer design*

QoE-based cross-layer optimisation is a topic being widely investigated. There are a number of studies that consider cross-layer optimisation for the sake of video quality enhancement, such as (Duong, Zepernick, & Fiedler, 2010; Gross, Klaue, Karl, & Wolisz, 2004; Gurses, Akar, & Akar, 2005) , or throughput improvement such as (Shabdanov, Mitran, & Rosenberg, 2012). We include only studies which are aimed at QoE improvement.

The Application/MAC/Physical (APP/MAC/PHY) cross-layer architecture introduced in (Khalek, Caramanis, & Heath, 2012) enables optimising perceptual quality for delay-constrained scalable video transmission. Using the acknowledgement (ACK) history and perceptual metrics, an online mapping of QoS to QoE has been proposed to quantify the packet loss visibility from each video layer. A link adaptation technique that uses QoS to QoE mapping has been developed at the PHY layer to provide perceptually-optimised unequal error protection for each video layer according to packet loss visibility. While at the APP layer, a buffer-aware source adaptation is proposed. The senders' rates are adapted by selecting a set of temporal and quality layers without incurring playback buffer starvation based on the aggregate channel statistics. To avoid frame re-buffering and freezing, a video layer-dependent per packet retransmission technique at the MAC layer limits the maximum number of packet retransmission based on the packet layer identifier. The next retransmission of packet is given a lower order of Modulation and Coding Scheme (MCS). The study concludes that the architecture prevents playback buffer starvation, handles short-term channel fluctuations, regulates the buffer size, and achieves a 30% increase in video capacity compared to throughput-optimal link adaptation. In addition to its limitation to SVC, the study didn't target a specific underlying wireless technology.

The QoE-driven seamless handoff scheme presented in (Politis, Dounis, & Dagiuklas, 2012) incorporates a rate adaptation scheme and the IEEE 802.21 Media Independent Handover (MIH) framework. The rate is controlled by adapting the Quantisation Parameter (QP) for the single layer coding (H.264/AVC) and dropping the enhancement layers for the scalable coding (H.264/SVC). The paper concluded that the proposed QoE-driven handover implemented in a real test-bed outperforms the typical Signal-to-Noise Ratio (SNR)-based handover and improves the perceived video quality significantly for both coding. However it can be better maintained with H.264/SVC. The study is merely a comparison between the two coding techniques for maintaining the QoE of wireless nodes during the handover process.

An online test-optimisation method is proposed in (Zhou, Yang, Wen, Wang, & Guizani, 2013) for resource allocation and optimisation of the total Mean Opinion Score (MOS) of all users without complete information of the QoE model (also called utility function of each user) or playout time (blind dynamic resource allocation scheme). Instead, MOS is observed over time dynamically. Each user subjectively rates the multimedia service given the allocated resource in the form of MOS value and reports it back to the base station. Dynamic resource allocation strategy learns user n's underlying QoE model by testing different allocated resources (testing) and seeks the optimal resource allocation solution (optimisation). The authors adopted the QoE prediction model in (A. Khan et al., 2010) for implementing the dynamic resource allocation scheme. The QoE model is estimated based on the observed MOS for the blind dynamic resource allocation scheme.

The application-driven objective function developed in (S. Khan, Peng, Steinbach, Sgroi, & Kellerer, 2006) optimises the quality of video streaming over the wireless protocol stack jointly by the application layer, data-link layer and physical layer. The proposed cross-layer optimiser periodically receives information in both directions, top-down and bottom-up from the video server and selects the optimal parameter settings of different layers. The optimisation is based on the outcome of maximisation of an object function which depends on the reconstruction quality in the application layer. The parameters that can be optimised are source rates at the application layer and modulation schemes, Binary Phase Shift Keying (BPSK) (total rate of 300kb/s) or Quaternary PSK (QPSK) (a total rate of 600 kb/s) in the radio link layer (radio link layer=physical + data link layer). The quality-based optimiser was applied to wireless users who simultaneously run voice communication, video streaming and file download applications in (Shoaib Khan, Duhovnikov, Steinbach, & Kellerer, 2007). QoE was measured in terms of Peak Signal-to-Noise Ratio (PSNR) and MOS mapped from an assumed linear PSNR to MOS mapping. It was assumed that a PSNR of 40 dB represents the maximum user satisfaction and 20 dB the minimum user satisfaction. It was compared to the conventional throughput optimiser and showed a significant improvement in terms of user perceived quality and wireless resource utilisation.

The application-driven cross-layer framework in (S. Khan et al., 2006) is extended to a QoE-based for High Speed Downlink Packet Access (HSDPA) (Thakolsri, Khan, Steinbach, & Kellerer, 2009). It combines both capabilities of HSDPA link adaptation and multimedia applications rate adaptation to maximise user satisfaction. Relevant parameters from radio link and application layers are communicated to a cross-layer optimiser. The optimiser acts as a downlink resource allocator and periodically reviews the total system resources and makes an estimate of the time-share needed for each user for each possible application-layer rate. It re-adapts the application rate if necessary. The QoE-based cross layer optimised scheme was simulated using OPNET against the throughput optimised & non-optimised HSDPA systems. It was concluded that user perceived quality significantly improved compared to the other two systems. The study made use of the adaptability feature of HTTP Adaptive Streaming (HAS) and aggressive TCP to control the application rate. Another shortcoming is that MOS was defined as a function of the transmission rate only.

Several techniques are proposed in (S. Latre, 2011) to optimise QoE in terms of the number of admitted sessions and video quality in the multimedia network. Traffic adaptation, admission control and rate adaptation are combined within an automatic management layer using both simulation and emulation on a large-scale testbed. The study focused on multimedia services such as IPTV and network-based personal video recording. Traffic flow adaptation modifies the network delivery of a traffic flow by determining required redundancy needed to cope with packet loss. An extension to the Pre-Congestion Notification (PCN)-based admission control system which is a distributed measurement based admission control mechanism recently standardised by the IETF. A novel metering algorithm based on sliding-window to cope with the bursty nature of video sessions and another adaptive algorithm to facilitate the configuration of PCN are proposed. Static and dynamic video rate adaptation algorithms that augment the PCN's binary-based (accept or reject) with the option of scaling video up or down. The viability of an implementation was investigated using neural networks and compared with an analytical model. The study shows that the QoE optimising techniques can successfully optimise QoE of multimedia services.

A generic and autonomic architecture presented in (S. Latre et al., 2009) to optimise the QoE of multimedia services. The proposed architecture is shown in Fig. 1. It comprises of Monitor, Action and Knowledge planes. The Monitor plane provides an automatic loop with a complete and detailed view of the network. Parameters such as packet loss, video frame rate and router queue size are monitored through monitor probes at demarcation points (e.g. access nodes, video servers). The Action plane optimises QoE based on a complete configuration of the actions received from the Knowledge plane. An example of these actions is adding the Forward Error Correction (FEC) packets to an existing stream after it has been determined by the Knowledge plane. The Knowledge plane based on the information from the Monitor plane and other relevant data such as historical information, detects network problems and bit errors on a link. It instructs the Action plane to take an appropriate QoE optimising action, e.g. switching to a lower bit rate video or adding an appropriate number of FEC packets. The Knowledge base component of the Knowledge plane stores relevant information about the network during each phase of the automation process (monitoring, reasoning and executing actions). The architecture was tested for optimisation of the QoE of video services in multimedia access networks using a neural network based reasoned. The reasoner applies FEC to reduce packet loss caused by errors on a link and switches to a different video bit rate to avoid congestion or obtain a better video quality. The authors concluded that their architecture was capable of increasing the video quality and lowering the packet loss ratio when packets are lost due to bit errors or when congestion occurs.

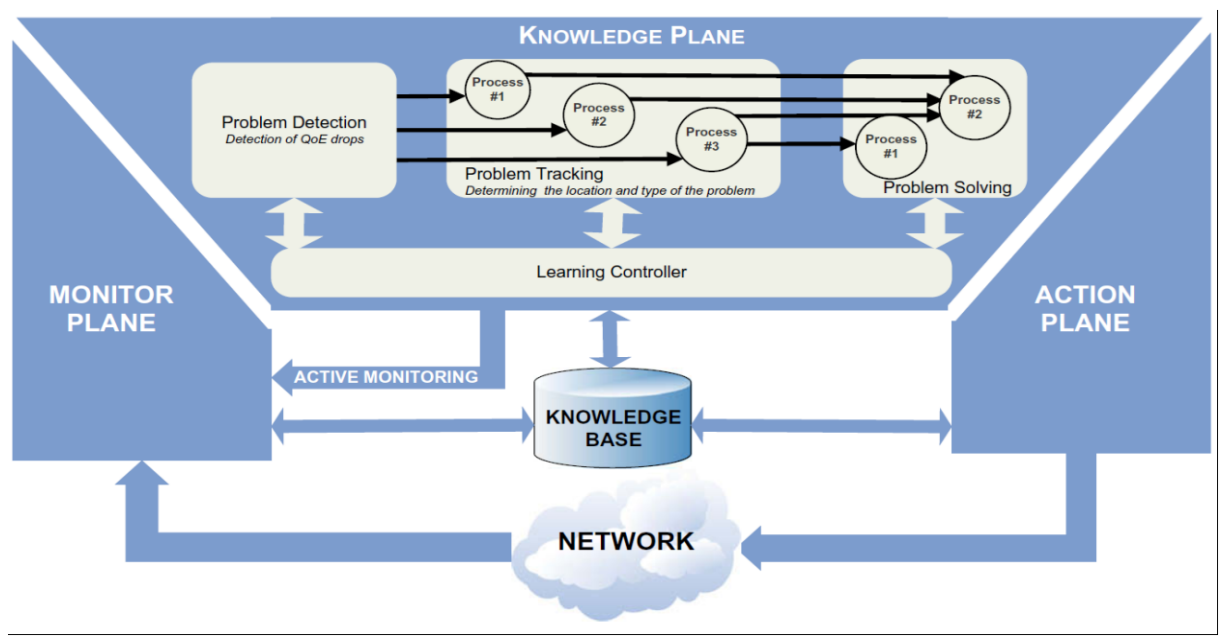

Figure 1. An automatic architecture to enable the QoE maximisation of multimedia services (S. Latre et al., 2009)

The cross-layer adaptation architecture shown in Fig. 2 is presented in (Oyman & Singh, 2012) for HAS-specific QoE optimisation. The layers of the architecture and corresponding layers of the Open Systems Interconnections (OSI) are depicted in the figure. It relies on tight integration of the HAS/HTTP-specific media delivery with network-level and radio-level adaptation as well as QoS mechanisms to provide highest possible end users' QoE. The following parameters are jointly involved between appropriate network layers:

1. Video level: bit rate, frame rate, resolution codecs
2. Transport level: Sequence and timing of HTTP requests, number of parallel TCP connections, HAS segment durations, frequency of Media Presentation Description (MDP) updates.

3. Radio and network level: Bandwidth allocation and multiuser scheduling, target QoS parameters for the core network and radio access network, MCS, Orthogonal Frequency Division Multiple Access (OFDMA) time/frequency resource/burst allocations.

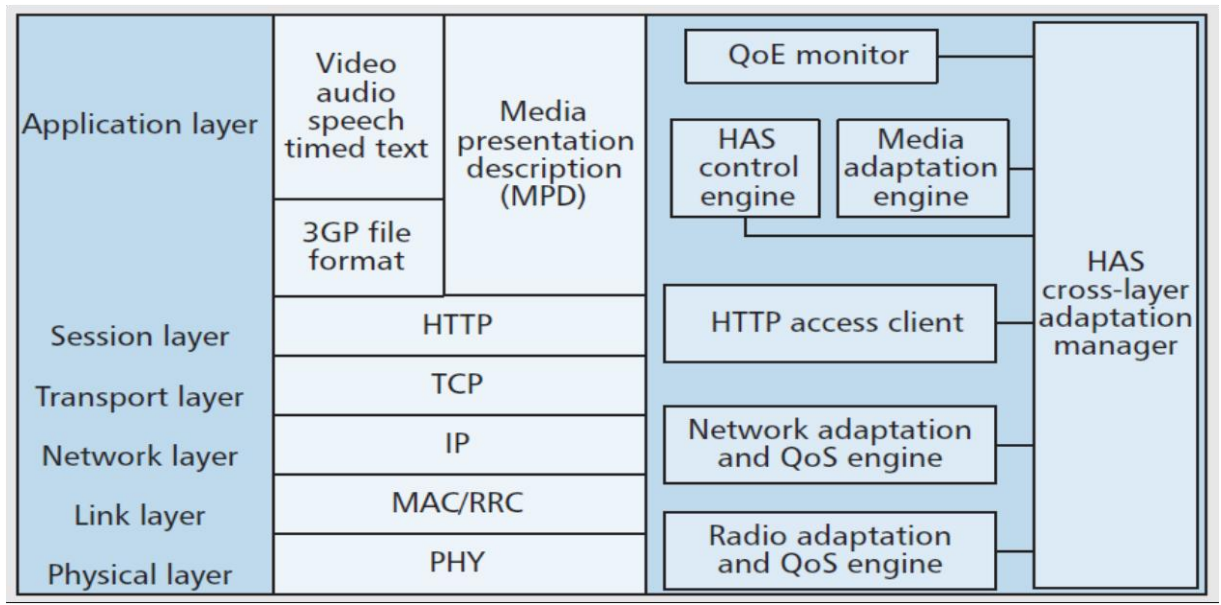

Figure 2. Cross-layer adaptation architecture for HAS-specific QoE optimisation (Oyman & Singh, 2012)

The end to end QoE optimisation system shown in Fig. 3 is proposed in (J. Zhang & Ansari, 2011) for Next Generation Network (NGN). The major elements of the QoE assurance framework as well as their functions are also depicted in the figure. The QoE/QoS reporting component at terminal equipment reports the user QoE/QoS parameters to the QoE management component. The transport functions and relevant parameters are analysed and adjusted accordingly. The updated QoS/QoE of end users is sent to the network and sources.

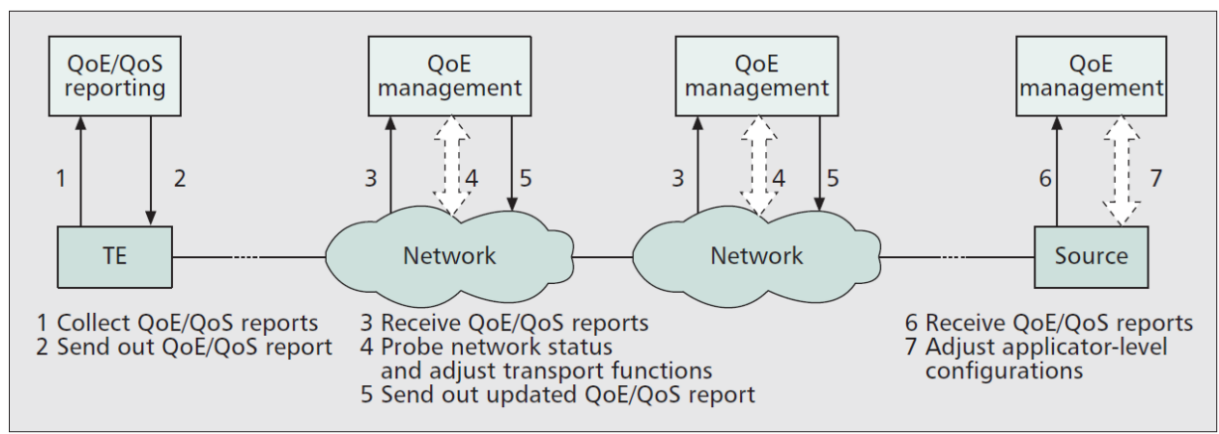

Figure 3. Possible end to end QoE assurance system (J. Zhang & Ansari, 2011)

A joint framework for video transport optimisation in the next generation cellular network is designed in (Fu et al., 2013). The rationale behind the design is to combine several optimisation approaches for more gain. As shown in Fig. 4, path selection, traffic management and fame filtering modules are proposed for SVC video streaming over UDP/RTP. The path selection module provides the best available end to end video path by redirecting the video traffic from a video source to another based on a set of network metrics. The traffic management module at the transport layer allocates transmission data rates for multiple video streams travelling through the core network nodes. The base station implements dynamic frame filtering to cope with the wireless channel variation. Issues such as WAN congestion, core network node congestion, cache failure and user mobility can be overcome by the presented design.

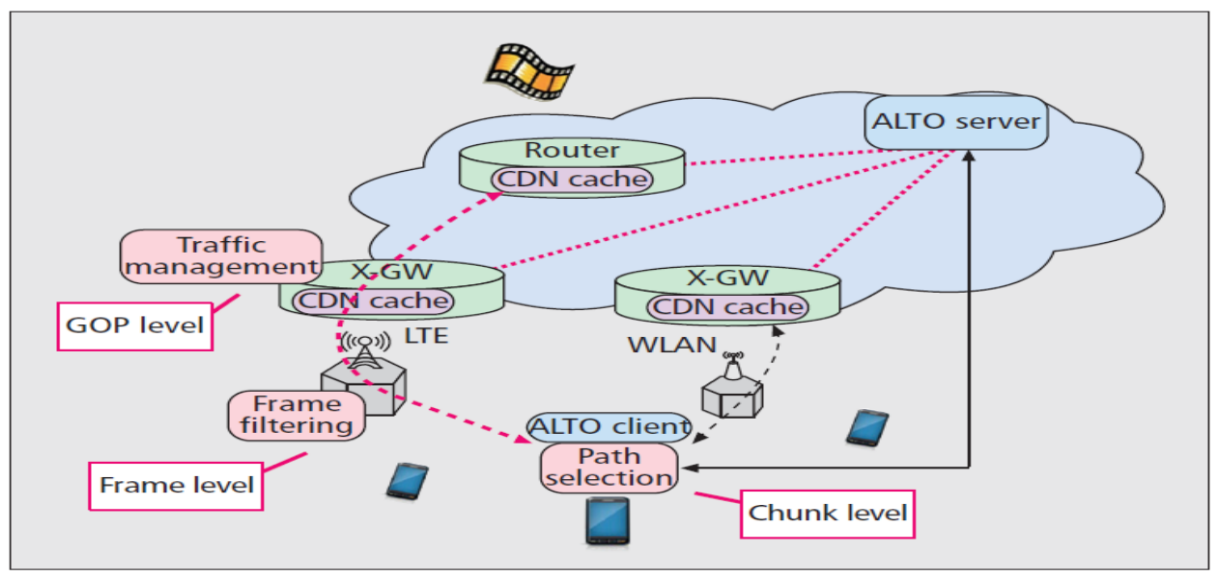

Figure 4. Joint framework for multilayer video optimisation (Fu et al., 2013)

HTTP-based Dynamic Adaptive Streaming (DASH) has attracted the attention of research community recently. A mobile DASH client decides on the streaming rate and the base station allocates resources accordingly. In contrast to the UDP push-based streaming, DASH is a pull-based client-driven streaming protocol (El Essaili, Schroeder, Steinbach, Staehle, & Shehada, 2014). The QoE-aware cross-layer DASH friendly scheduler introduced in (Zhao et al., 2014) allocates the wireless resources for each DASH user. The video quality is optimised based on the collected DASH information. Furthermore, an improved SVC to DASH layer mapping is proposed to merge small sized layers and decrease overhead. For smooth playback, along with the existing client-based quality selection policies, there is a DASH proxy-based which transparently stabilises bitrates. The authors concluded that their proposed scheme outperforms others schemes. A proactive approach for optimising multi-user adaptive HTTP video QoE in mobile networks is proposed in (El Essaili et al., 2014). In contrast to the reactive approach in which resources are allocated by the mobile operator without clients' knowledge, in the proactive approach a proxy overwrites the client HTTP request based on the feedback from a QoE optimiser. The QoE optimiser on the base station collects information about each client and determines the transmission rate and signals it back to the proxy and resource shaper for adapting the transmission rate of DASH client. The proxy ensures the streaming rate is supported by lower layers and QoE optimisation. Subjective test is conducted for end users' perception on QoE.

Two QoE-aware joint subcarrier and power radio resource allocation algorithms are presented in (Rugelj et al., 2014) for the downlink of a heterogeneous OFDMA system. They allocate resources based on the QoE of each heterogeneous service flow. A utility function maximising the minimum MOS experienced by users considered by the first algorithm and the second algorithm balances between the level of QoE and system spectral efficiency. Each user of the OFDMA system can achieve an appropriate level of QoE through adaptable resource allocation and data rate. Numerical simulation results presented a significant increase of QoE achieved by the algorithms compared to the data rate maximisation-based algorithms.

A joint near optimal cross-layer power allocation and QoE maximisation scheme for transmitting SVC video over the Multi-Input Multi-Output (MIMO) systems proposed in (Chen, Hwang, Lee, & Chen, 2014). The effect of power allocation to bit error rate in the physical layer and video source coding structures in the application layer are considered. The scheme is further extended with the Reed-Solomon (RS) code and different MCS. The calculated PSNR and Structural SIMilarity (SSIM) from simulation demonstrated the efficiency of the scheme over the water-filling (WF) and modified-WF schemes.

An application-level signalling and end-to-end negotiation called Media Degradation Path (MDP) is deployed in (Ivesic, Skorin-Kapov, & Matijasevic, 2014) for resource management of the adaptive multimedia services in Long-Term Evolution (LTE). Admission control and resource reallocation in case of limited resource availability as two components of the cross-layer design increase session admission rate while maintaining an acceptable level of end users' QoE. Alternative configuration of MDP is applied to a new session if the available resources are not sufficient for the optimal configuration. Since, both configurations are set with users' preference and acceptable quality level, users' satisfaction are kept at an acceptable level. The authors considered the impact on end users' QoE from the perspective of performing utility-driven adaptation decisions, improving session establishment success, and meeting QoS requirements (i.e. loss thresholds). Neither subjective nor objective MOS is taken into account.

Work in (Debono et al., 2012) address the issue of high delay computational power caused by video error concealment techniques at receivers. The QoE of the region of a mobile physician's interest is optimised by adopting a cross-layer design approach in mobile worldwide interoperability for microwave access wireless communication environment while ensuring real-time delivery. Advanced concealment techniques are applied if the Region Of Interest (ROI) is affected and a standard spatial or temporal concealment if it is otherwise. The cross-layer parameters are determined to reduce the packet error rates by utilising the QoE of the ROI. The strategy does not demand a higher bandwidth as the quality is optimised through better error concealment not encoding with a higher QP. A PSNR of about 36 dB was obtained within reasonable decoding time.

Work presented in (Singhal, De, Trestian, & Muntean, 2014) combines various techniques across different layers for optimisation of both users' QoE levels and energy efficiency of wireless multimedia broadcast receivers with varying display and energy constraints. The SVC optimisation, optimum time slicing for layer coded transmission, and a cross-layer adaptive MCS are combined to present a cross-layer framework. Users are grouped based on their device capability and channel condition and they are offered options to trade between quality and energy consumption. The scheme compared to energy saving based optimisation, achieved a 43% higher video quality trading off 8% in energy saving and a marginal 0.62% in user serving capacity, whereas compared to quality based optimisation, the scheme results in 17% extra energy saving, 3.5% higher quality, and 10.8% higher capacity.

Work in (Mathieu et al., 2011) argues that the end-to-end QoE can be improved by advocating a close cooperation between ISPs and applications via a comprehensive, media-aware and open Collaboration Interface between Network and Applications (CINA). Mutual information is exchanged between the network layer and applications through CINA which bridges the two entities. CINA and other components to support this cooperation are shown in Fig. 5. The system is expected to support service providers to efficiently distribute highly demanding content streams and enable dynamic adaptation to satisfy the requirement of users within the underlying network capability. The internal functionality of each block and evaluation through both simulation and testbed are identified as future work.

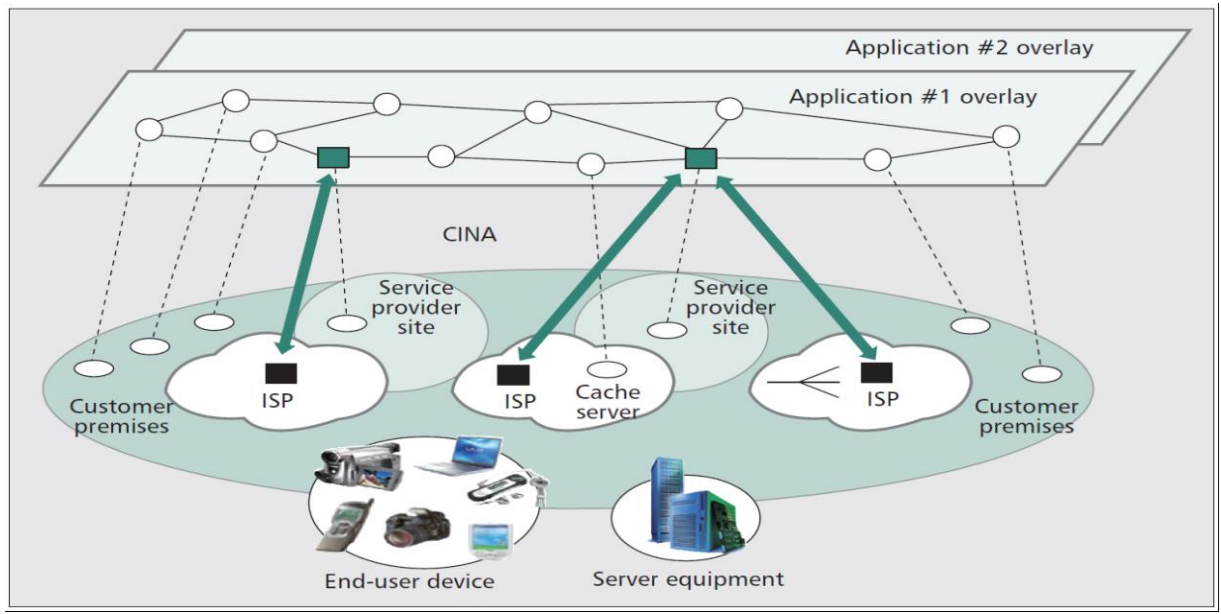

Figure 5. Overview of the cooperation system components and their relationships (Mathieu et al., 2011)

In (Pejman Goudarzi, 2012) particle swarm optimisation is utilised to find an optimal rate by which the total weighted QoE of some competing video sources is optimised. It is also used for differentiated QoE enforcement between multiple competing scalable video sources. Scalable video encoder such as H.264/MPEG4 AVC can use the resulting rate for an online rate adaptation. The work presented in (P. Goudarzi & Hosseinpour, 2010) adopts a model from literature to capture the exact effect of network packet loss and finds the optimal rate toward minimising the loss-induced distortion associated with video sources and maximising QoE. The resulting optimal rate is sent back to video encoders for the online rate adaptation.

A cross-layer scheme for optimising resource allocation and user perceived quality of video applications based on a QoE prediction model that maps between object parameters and subject perceived quality is presented in (Ju, Lu, Zheng, Wen, & Ling, 2012). Work presented in (Fiedler, Zepernick, Lundberg, Arlos, & Pettersson, 2009) promotes automatic feedback of end-to-end QoE to the service level management for better service quality and resource utilisation. A QoE-based cross-layer design of mobile video systems is presented for this purpose. Challenges of incorporating the QoE concepts among different layers and suggested approaches span across layers such as efficient video processing and advanced realtime scheduling are also discussed.

In (Qadir, Kist, & Zhang, 2014) the issue of QoE degradation of video traffic in a bottleneck network is addressed by introducing a QoE-aware cross-layer architecture for optimising the video quality shown in Fig. 6. In particular, it allows video sources at the application layer to adapt themselves to the network environment by controlling their transmitted bit rate dynamically, and the edge of network to protect the quality of active video sessions at the network layer by controlling the acceptance of new session through a QoE-aware admission control. The application layer contributes to the optimisation process by dynamically adapting source bit rate based on the condition of network and the network layer controls admission of new video session based on a QoE-aware admission control.

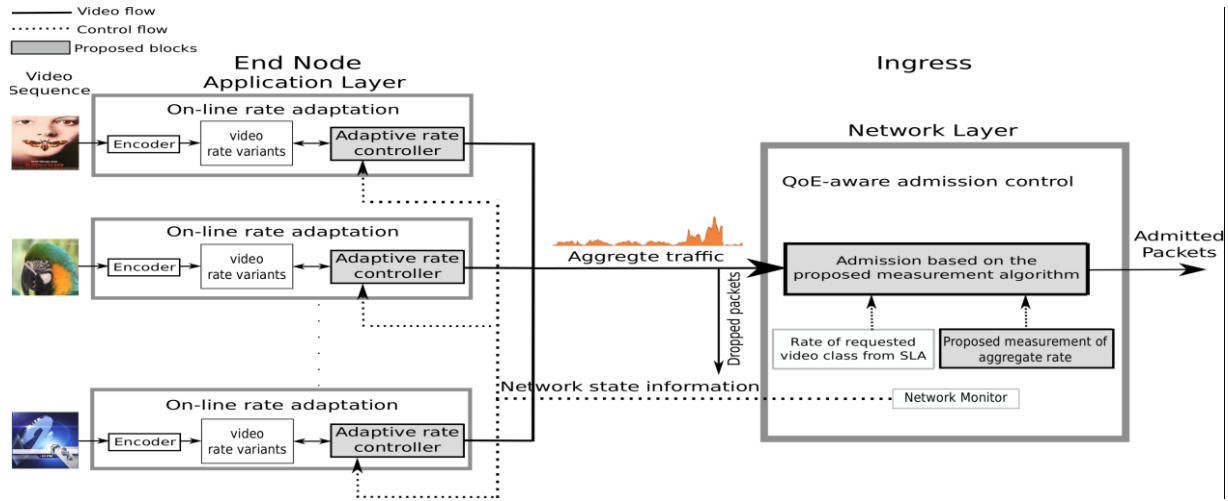

Figure 6. QoE-aware cross-layer architecture for video traffic (Qadir et al., 2014)

A comparison of mechanisms relying on cross-layer design for QoE optimisation is shown in Table II.

## C. QoE optimisation through scheduling

In contrast to scheduling strategies based on QoS metrics such as delay, jitter or packet loss, QoE-aware schedulers have been proposed by researchers. Individual user's QoE is included in a QoE-aware scheduler through one-bit feedback from user to indicate their satisfaction (Lee, Kim, Cho, & Lee, 2014). The derived user-centric QoE function modelled by the Sigmoid function can significantly improve the average QoE and fairness for wireless users.

The packet scheduler presented in (Navarro-Ortiz et al., 2013) improves the QoE of HTTP video users that prioritises flows based on the estimation of the amount of data stored in the players' buffer. Simulation results showed a reduction of the number of pauses at receivers' video playback for OFDMA based system such as 3G LTE and IEEE 802.16e.

Work in (Taboada, Liberal, Fajardo, & Ayesta, 2013) focuses on the delay as a main distortion factor over others such as packet loss ratio. A delay-driven QoE-aware scheduling scheme is proposed based on the Markov decision process. Gittins index rule was developed for the scheme which gives the priority to flows that are statistically closer to finish and those whose QoE has not been degraded too much. The rule is a combination of the attained service-dependent completion probability and delay-dependent MOS function. Compared to Round Robin, FIFO and Random, the scheduler outperforms in terms of delay and MOS.

A comparison of mechanisms relying on scheduling for QoE optimisation is illustrated in Table III.

Table II. COMPARISON OF CROSS-LAYER MECHANISMS

| Ref. | Approach | Traffic | Date | Underlying network | QoE measurement | Limitations |
|---|---|---|---|---|---|---|
| (S. Latre, 2011) | PCN-based admission control, rate adaptation, redundancy | Video | 2011 | Multimedia access Net. | PSNR, SSIM | Missing subjective MOS |
| (Khalek et al., 2012) | Link adaptation, buffer-aware rate adaptation, layer-dependent retransmission | Video | 2012 | Wireless | MS-SSIM | Limited to SVC |
| (Chen et al., 2014) | Transmission error & video source coding characteristic | SVC Video | 2014 | MIMO System | PSNR & SSIM | Missing subjective MOS |
| (Ivesic et al., 2014) | Admission control & resource reallocation | Adaptive multimedia service | 2014 | 3GPP & LTE | Session establishment success, meeting QoS requirement | QoE not measured objectively or subjectively |
| (Debono et al., 2012) | Coding, FEC, ARQ, modulation coding | Ultrasound video | 2012 | Mobile WiMAX | PSNR | Missing subjective MOS |
| (Singhal et al., 2014) | SVC optimisation, cross-layer MCS, optimum time | QCIF, CIF, D1 | 2014 | Wireless | Utility function dependent on QP & frame rate | Missing subjective MOS |
| (Shoaib Khan et al., 2007) | Cross-layer optimiser | QCIF | 2007 | Wireless | PSNR & MOS | MOS mapped from PSNR |
| (S. Khan et al., 2006) | Source rate adaptation, estimate wireless capability & Quickly adapting to its variation | QCIF | 2006 | Wireless | PSNR | MOS mapped from PSNR |
| (Mathieu et al., 2011) | Overview block design | Not specified | 2011 | Not specified | None | Missing evaluation |
| (Qadir et al., 2014) | QoE-aware admission control, rate adaptation | QCIF | 2014 | Internet | PSNR & MOS | QoE mapped from PSNR |
| (J. Zhang & Ansari, 2011) | QoE assurance framework | Video | 2011 | NGN | None | Missing evaluation |
| (Politis et al., 2012) | MIH, QoE-driven rate adaptation | Video | 2012 | WiFi, 3G/UMTS | PSNR & Subjective MOS | None |
| (Zhou et al., 2013) | Dynamic resource allocation | QCIF, audio | 2013 | Wireless | Subjective MOS | Non-dynamic QoE model |
| (Fu et al., 2013) | Joint framework | Video | 2013 | Cellular | Utility function dependent on delay | QoE estimated from delay only |
| (Zhao et al., 2014) | SVC-DASH mapping, DASH friendly scheduler, resource allocation, DASH-based proxy rate stabiliser | Streaming video over HTTP | 2014 | Wireless broadband access | Average PSNR | QoE mapped from PSNR |
| (El Essaili et al., 2014) | QoE-based traffic & resource management | Video | 2014 | LTE | Subjective MOS | Buffer level-based QoE optimisation considered instead of stream-based optimisation |
| (Rugelj et al., 2014) | Radio resource allocation | Video, audio, best-effort | 2014 | OFDMA | Utility function given by Eq. 8 in the literature | QoE not measured objectively or subjectively |
| (S. Latre et al., 2009) | Adding redundancy, video adaptation | Video | 2009 | Multimedia access | SSIM, PSNR | Missing subjective MOS |
| (Oyman & Singh, 2012) | Network and radio levels adaptation, QoS mechanisms | Video streaming | 2012 | 3GPP LTE | None | Missing evaluation |
| (Thakolsri et al., 2009) | HSDPA link adaptation, multimedia application rate adaptation | Video | 2009 | HSDPA | MOS adopted utility function dependent on transmission rate & packet loss rate, SSIM | Missing subjective MOS |
| (Pejman Goudarzi, 2012) | Optimum rate found by swarm algorithm | Video | 2012 | Wireless | Adopted utility function (Eq. 7 in the literature) | QoE not measured objectively or subjectively |
| (P. Goudarzi & Hosseinpour, 2010) | Optimum rate found by an adopted model (Eq. 9 in the literature) | Mobile video | 2010 | MANET | PSNR-MOS mapping of (S. Khan et al., 2006) | QoE mapped from PSNR |

Table III. COMPARISON OF SCHEDULING MECHANISMS

| Ref. | Approach | Traffic | Date | Underlying network | QoE measurement | Limitations |
|---|---|---|---|---|---|---|
| (Lee et al., 2014) | QoE-aware scheduling | Mobile video | 2014 | Wireless | Utility function given by Eq. 10 in the literature | Missing evaluation |
| (Navarro-Ortiz et al., 2013) | Packet scheduling | Mobile video streaming | 2013 | Wireless | Number of playback interruption | QoE estimated based-on the reduction of playback interruption |
| (Taboada et al., 2013) | Delay-driven QoE-aware scheduling | Video | 2013 | Wireless | Utility function dependent on delay | QoE model based-on delay only |

## D. QoE optimisation through content and resource management

Buffer starvation is analysed through two proposed approaches in (Xu et al., 2014) to obtain exact distribution of the number of starvations. They are applied for QoE optimisation of media streaming. The first approach is based on Ballot theorem and the second uses recursive equations. The fluid analysis-based starvation behaviour controls the probability of starvation on the file level. Subjective human unhappiness is modelled using an objective QoE cost which is a weighted sum function of the start-up/rebuffering delay and starvation behaviour. They are taken as quality metrics as the QoE of streaming service is considerably affected by them. The weight reflects an individual user's relative impatience on the delay rather than starvation.

A content cache management for HTTP Adaptive Bit Rate (ABR) streaming over wireless networks and a logarithmic QoE model from experimental results are formulated in (W. Zhang, Wen, Chen, & Khisti, 2013). Alternative search algorithms to find and compare the optimal number of cached files are also provided. The numerical results suggested high QoE with low complexity can be provided under the optimal cache schemes.

Work in (Latre, S., Roobroeck, Wauters, & Turck, 2011) presents an extended architecture of the PCN-based admission control to protect video services. Three modifications (highlighted block) are proposed to the original PCN systems as shown in Fig. 7. First, the sliding-window-based bandwidth metering algorithm instead of the traditional token bucket finds the highest rate value that avoids any congested related losses. Second, to reduce the required headroom, packets are buffered just before the PCN metering function. Third, a video rate adaptation algorithm decides on each video quality level based on the current network load. The performance of the modified PCN architecture was evaluated and resulted in an increase of 17% in the network utilisation for the same video quality.

Content encoding for video streaming is addressed with the aim of reducing bitrates and optimising QoE in (Adzic, Kalva, & Furht, 2012). A process for content-based segmentation from the encoding stage to segmentation stage is proposed for the adaptive streaming over HTTP. It can tailor video streams with better QoE while saving 10% of the bandwidth on average for the same quality level.

Changing between mobile-television programs is called zapping which is not immediate but there is a finite delay called zapping delay. A known bound of zapping-delay in Digital Video Broadcast-Handheld (DVB)-H is found in (Vadakital & Gabbouj, 2011) as a way to maximise the QoE of mobile video services. Video prediction structures and their reception in time-sliced bursts are analysed using graph theoretic principles. The authors concluded that their system guarantees a zapping delay below some maximum threshold and gradually enhances the quality of video after zapping.

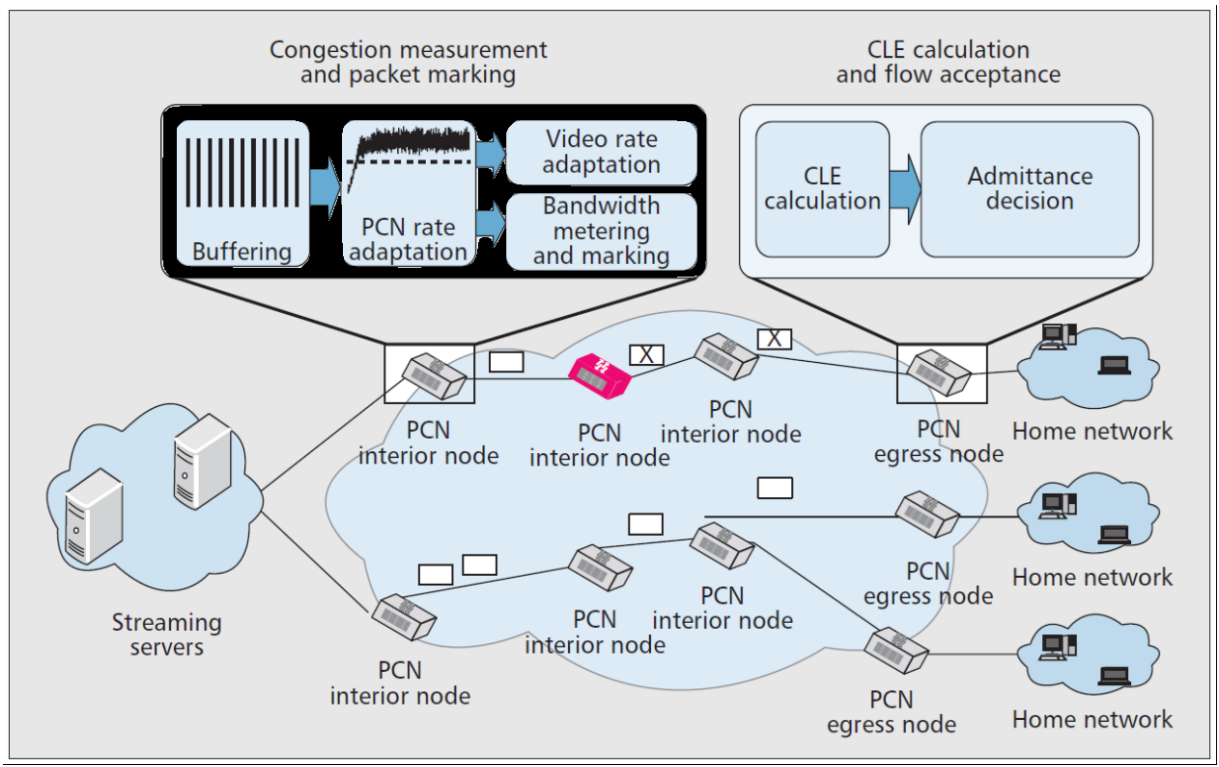

Figure 7. Modification of the PCN-based admission control system toward the optimisation of video services in access network (Latre et al., 2011)

A comparison of mechanisms relying on managing content and resource for QoE optimisation is illustrated in Table IV.

Table IV. COMPARISON OF CONTENT AND RESOURCE MANAGEMENT MECHANISMS

| Ref. | Approach | Traffic | Date | Underlying network | QoE measurement | Limitations |
|---|---|---|---|---|---|---|
| (Xu et al., 2014) | Buffer starvation analysis | Video on demand | 2014 | Not specified | Objective QoE cost | Missing evaluation |
| (Latre et al., 2011) | Bandwidth metering, buffering, video rate adaptation at router | Streaming video | 2011 | Multimedia access | SSIM, session, utilisation | Missing subjective MOS |
| (W. Zhang et al., 2013) | Content cache management | HTTP ABR streaming | 2013 | Wireless | Utility function dependent On required & actual playback rate-based | Non-uniform distribution request & multiple distinctive content on cache not considered |
| (Adzic et al., 2012) | Content-based segmentation, optimised content preparation algorithm, encoding | Adaptive streaming video | 2012 | Not specified | PSNR | QoE estimated from PSNR |
| (Vadakital & Gabbouj, 2011) | Bounding Zapping-delay | Video | 2011 | DVB-H | Zapping delay-dependent | Zapping-event between two bursts not considered |

## IV. OPEN ISSUES

The first step of QoE optimisation is to measure QoE in an accurate way. Current QoE estimation models are limited to specific video resolutions and coding schemes. Thus, finding a prediction model that can estimate the quality for as wide as possible of different video formats and coding is required. As per the recommendation of ITU, any attempt for QoE modelling has to consider objective modelling of measurable technical performance and subjective testing with people (Brooks & Hestnes, 2010). More intelligence fairness techniques are useful to avoid penalising the same user in case of insufficient resources where some traffic needed to be dropped. Cross-layer designs have to consider more relevant parameters to achieve better optimised outcome.

## V. CONCLUSION

This paper has surveyed studies have been done in the area of QoE optimisation for video traffic in the last 10 years. Challenges in achieving optimised video quality and motivation for this objective have been discussed in details. They have been classified into groups based on their functions; rate-adaptation, cross-layer architecture, scheduling, content and resource management. The limitation of each of these studies has been identified and future potential research areas have been highlighted.